\documentclass[aps,prd,onecolumn,showpacs,nofootinbib,superscriptaddress, 12pt]{revtex4-2}
\usepackage{amsmath,amssymb,amsfonts, bm}
\usepackage{graphicx}
\usepackage{subcaption}
\usepackage{hyperref}

\begin{document}
\title{Gauss-Bonnet Gravity and Spacetime Singularities}

\author{Tariq Allaithy}
\email{tariqallaithy@mans.edu.eg}
\affiliation{Department of Physics, Faculty of Science, Mansoura University,\\ Mansoura 35516, Egypt}

\author{Adel Awad}
\email{a.awad@sci.asu.edu.eg}
\affiliation{Department of Physics, Faculty of Science, Ain Shams University,\\ Cairo 11566, Egypt}
\affiliation{Centre for Theoretical Physics, British University of Egypt,\\
Sherouk City 11837, Egypt}

\author{Mohamed Hany Radwan}
\email{m.radwan@uky.edu}
\affiliation{Department of Physics and Astronomy, University of Kentucky, \\ Lexington, KY 40506, U.S.A.}

\author{Mohsen Zahran}
\email{m_zahran1@mans.edu.eg}
\affiliation{Department of Physics, Faculty of Science, Mansoura University,\\ Mansoura 35516, Egypt}

\maketitle

\section*{Abstract}
We investigate the effect of higher-order curvature terms, specifically Gauss–Bonnet terms, on spacetime singularities in five dimensions. For FLRW cosmologies, we demonstrate that Gauss–Bonnet terms can replace the Big Bang/Crunch with a "sudden" singularity, characterized by a finite scale factor and Hubble rate but diverging higher-order derivatives. Investigating various branches of solutions shows the possibility of explicit extension of non-spacelike geodesics beyond the singular point. Furthermore, we employ the Gauss–Bonnet junction conditions to verify the consistency of the extension with the field equations. The whole solution describes a contracting phase prior to the expansion phase with a well-defined surface stress–energy tensor. Regarding the Boulware–Deser black hole, we find that Gauss–Bonnet terms soften the central singularity for radial geodesics—rendering them "weak" according to the Tipler and Królak criteria—whereas non-radial geodesics remain strongly singular. Junction condition analysis of this solution shows that although higher-curvature corrections alter the nature of the singularity, geodesics are still inextendible as a result of divergent extrinsic curvature. Our results are consistent with the Penrose-Hawking singularity theorems since in Gauss-Bonnet black holes, geodesics suffer from focusing (expansion parameter diverges), while in cosmology, there is no focusing since the expansion parameter remains finite at the singularity.

\section{Introduction}

While General Relativity gives a very successful description of gravity, it is expected to be modified at high-curvature scales, such as those near a singularity. Naturally, this could lead us to consider $f(R)$ theories of gravity, but these theories suffer from Ostrogradsky instabilities and ghost degrees of freedom \cite{Gao_2012, Chen_2013}. This is why many authors consider working with more controlled theories such as Lovelock gravity, since it avoids these problems \cite{ZWIEBACH1985315, ZUMINO1986109}.\\

Lovelock gravity in $D$ dimensions \cite{Lovelock:1971yv, Lovelock:1970zsf} is a metric theory of gravity which has the most general Lagrangian density that leads to second-order field equations. Its Lagrangian is a linear combination of Euler densities. Within Lovelock theory, the zeroth-order term represents the cosmological constant, while the first-order term corresponds to the Einstein–Hilbert action ($R$). The second-order contribution is the Gauss–Bonnet (GB) invariant, given by the quadratic curvature combination $\mathcal{L}_{GB} = R^2 - 4R_{\mu\nu}R^{\mu\nu} + R_{\mu\nu\rho\sigma}R^{\mu\nu\rho\sigma}$. In four dimensions of spacetime ($D=4$), the Gauss–Bonnet term is purely topological, equal to the Euler characteristic, and contributes only a total derivative to the action, so it does not affect the local dynamics. In contrast, for five or more dimensions ($D \geq 5$), the Gauss–Bonnet term becomes nontrivial, providing quadratic curvature corrections that modify the behavior of the theory.\\

The Einstein-Gauss-Bonnet formulation can be motivated from a low-energy Effective Field Theory (EFT) perspective. In an EFT expansion of gravity, the standard Einstein-Hilbert action is supplemented by an infinite series of higher-curvature corrections with a characteristic length scale $L$. To leading order in this expansion, the action takes the generic quadratic form $S \sim \int d^5x \sqrt{-g}\frac{1}{2\kappa} [R + L^2 (c_1 R^2 + c_2 R_{\mu\nu}R^{\mu\nu} + c_3 R_{\mu\nu\rho\sigma}R^{\mu\nu\rho\sigma}) + \mathcal{O}(L^4)]$. However, for arbitrary dimensionless coefficients $c_i$, the variation of these terms generically produces fourth-order field equations, leading to Ostrogradsky instability \cite{Kaparulin_2014, Smilga_2017}. To represent a physically stable theory, we ask that the phase space does not acquire ghost degrees of freedom. This mandates the specific structure of the expansion coefficients to $c_1 = c_3 = -c_2/4 \equiv \alpha$ \cite{PhysRevLett.55.2656, Donoghue_1994}. This uniquely isolates the Gauss-Bonnet term, making it the most general diffeomorphism-invariant theory of gravity that yields strictly second-order equations of motion.\\

This makes Einstein–Gauss–Bonnet (EGB) gravity deeply rooted in string theory, since in heterotic string theory, the low-energy effective action naturally incorporates quadratic curvature corrections, which can be put in Gauss-Bonnet. These corrections come with the coupling,  $\alpha$, which is identified as the inverse string length \cite{KUIROUKIDIS_2006}, with $\alpha > 0$ (for example, see \cite{dey2022anisotropicstrangestarseinstein}), and in the context of low-energy effective field theory, the framework used in this paper is valid in the regime where $\alpha/L^2 \ll 1$, with $L$ being the characteristic spacetime curvature scale. This is why many authors regard 5D EGB gravity as a good ground for exploring higher-curvature effects in general relativity, yielding a broad spectrum of phenomena, from modified black hole thermodynamics \cite{Fairoos_2024,PhysRevD.100.024001,Cvetic:2001bk} to modified cosmological solutions\cite{Tsujikawa_2004,Deruelle:1989fj,DERUELLE198725}.\\

One of the consequences of incorporating these corrections is the softening of the FLRW Big Bang/Crunch singularity into a sudden singularity. The latter singularity was introduced in \cite{Barrow__2004, Barrow_2004, BARROW_2012} when the authors studied general features of the FLRW cosmologies in certain contexts. A sudden singularity is defined by the divergence of the second derivative of the scale factor, $\ddot{a}$, while both $a$ and $\dot{a}$ remain finite. Interest in such singularities has surged following the discovery of cosmic acceleration \cite{Riess_1998, Riess_2004, Knop_2003, Perlmutter_1999, Bennett_2003, Spergel_2003}, as they emerge naturally within various cosmological models, such as those present in \cite{PhysRevD.85.124012, Shtanov_2002, Tretyakov_2006, PhysRevD.86.063522, Kamenshchik_2001, Bili__2002, Bento_2002, Nojiri_2004,Bamba_2008,Bamba2010}. The 5D EGB cosmology can provide us with mechanisms explaining cosmic acceleration and early-universe singularities. In this setting, the modified Friedmann equations include quartic powers of the Hubble parameter, producing new dynamical behaviors. For a 5D Friedmann-Lemaître-Robertson-Walker (FLRW) geometry, the corresponding field equations lead to a modified expansion history. The Friedmann equation is modified to $H^2 + \alpha H^4 \sim \rho$, where $\rho$ is the energy density \cite{Deruelle:1989fj,DERUELLE198725}. At high energies (in the early universe), the $H^4$ contribution dominates, leading to the Hubble parameter $H \propto \rho^{1/4}$ instead of the usual $H \propto \rho^{1/2}$. This weaker dependence on $\rho$ helps in avoiding the initial singularity by modifying the Raychaudhuri equation; the Gauss–Bonnet term effectively produces a repulsive force that counteracts gravitational focusing \cite{G_mez_2022}. As a result, the Big Bang singularity characteristic of standard GR cosmology can be replaced by a sudden singularity in 5D EGB gravity. In such a scenario, the universe undergoes a phase of contraction down to a finite minimum scale factor $a_{\text{min}}$ before transitioning into a re-expanding phase \cite{Madhunlall_2025}. An important feature of a spacetime with a sudden singularity is that it is transversal, since its geodesics are well-behaved and can be extended through the singular region \cite{Fernandez-Jambrina:2006qld, PhysRevD.74.064030, Barrow_2013}. This is illustrated by Keresztes et al. \cite{PhysRevD.86.063522} using an anti-Chaplygin gas and dust model. Moreover, the findings in this paper highlight some sort of universal correction, due to the $H^4$ term, which also appears in \cite{Awad_2016, Awad_2017}, where the first considers the FLRW flat cosmology under the Weyl anomaly corrections, leading to the same quartic term in $H$! The latter work considers generalized teleparallel gravity with $f(T) = T + \tilde{\alpha}(-T)^n$ where the higher-torsion terms intriguingly lead again to the $H^4$ term.\\

Furthermore, in this work, we are going to investigate the singularity of Boulware and Deser solution, which is a static spherically symmetric solution in Gauss-Bonnet gravity. This solution represents an extension of the Schwarzschild–Tangherlini metric that incorporates quadratic curvature contributions. The metric function $f(r)$ appears in the line element, which is determined by solving the polynomial equation that follows from the field equations. A key property of this solution is that it admits two separate branches: the (-) branch, which smoothly approaches the Schwarzschild–Tangherlini solution in the limit $\alpha \to 0$, and the (+) branch, which is unstable and regarded as unphysical \cite{PhysRevLett.55.2656, GARRAFFO_2008}.\\

For the "minus" branch, the gravitational potential is altered at short distances. Although the large-distance behavior still agrees with Newtonian gravity in five dimensions, the Gauss–Bonnet term introduces a repulsive component at small scales. This repulsion effectively softens the central singularity, even though it does not completely remove it in the vacuum case.\\

In short, our aim here is to investigate how higher-order curvature corrections in Einstein–Gauss–Bonnet (EGB) gravity modify the nature, strength, and geodesic structure of spacetime singularities in five dimensions, both in cosmological settings and in black hole spacetimes. 

The rest of the paper is organized as follows: In sec. II, we provide the Gauss-Bonnet setup, in Sec. III, we focus on the initial cosmological singularity in a 5D FLRW background: Sec. III A derives the modified Friedmann equations in EGB gravity, Sec. III B introduces a mechanical analog to interpret the dynamics and the emergence of a sudden singularity, Sec. III C constructs a $C^2$ extension of non-spacelike geodesics across this singular point and analyzes its strength using the Tipler–Królak criteria, and Sec. III D applies Gauss–Bonnet junction conditions to glue the two cosmological branches into a geodesically complete bouncing spacetime with a well-defined surface stress–energy, in Sec. IV, we turn to the black hole singularity, studying timelike and null geodesics in the Boulware–Deser solution; we show that while non-radial geodesics remain strongly singular due to centrifugal effects, purely radial geodesics encounter only a weak singularity according to the Tipler–Królak integrals and can be extended through $r=0$, in Sec. V, we check our results against the Hawking–Penrose singularity theorems, finally, in Sec. VI, we give our conclusion. 

\newpage
\section{The Gauss-Bonnet setup}

The general action of the EGB gravity can be expressed as
\begin{equation}
S = \int d^5x \, \sqrt{-g}  \left[\frac{1}{2\kappa} \left(R + \alpha \mathcal{L}_{GB} \right)+ \mathcal{L}_m \right]
\end{equation}

where \(R\) is the Ricci scalar, \(\mathcal{L}_m\) denotes the matter Lagrangian, and \(\alpha\) is the Gauss–Bonnet coupling constant with dimensions of \([{\rm L}]^2\). The Gauss–Bonnet term is given by

\begin{equation}
\mathcal{L}_{GB} = R^2 - 4R_{\mu\nu}R^{\mu\nu} + R_{\mu\nu\rho\sigma}R^{\mu\nu\rho\sigma}
\end{equation}

Varying the action with respect to the metric tensor \(g_{\mu\nu}\) gets us the modified field equations,

\begin{equation}
G_{\mu\nu} + \alpha H_{\mu\nu} = \kappa T_{\mu\nu}
\end{equation}

where \(G_{\mu\nu}\) is the Einstein tensor, and \(H_{\mu\nu}\) is the Gauss–Bonnet correction tensor defined as

\begin{equation}
H_{\mu\nu} = 2\left(
R R_{\mu\nu} - 2R_{\mu\rho}R^{\rho}{}_{\nu} - 2R^{\rho\sigma}R_{\mu\rho\nu\sigma} + R_{\mu}{}^{\rho\sigma\lambda}R_{\nu\rho\sigma\lambda} \right) - \frac{1}{2}g_{\mu\nu}\mathcal{L}_{GB}.
\end{equation}

The Gauss-Bonnet term is the simplest curvature correction to General Relativity. While in four dimensions this term is a topological invariant and thus has no impact on spacetime dynamics, in five or more dimensions it becomes dynamic. It directly influences the spacetime curvature, especially at high energies. Thus, EGB theory provides the perfect setting within which to study what kinds of effects high-curvature corrections may have on the behavior of spacetime singularities.\\


\section{The Initial Singularity}
\subsection{The GB FLRW Cosmology}
We start by looking at the initial singularity in the FLRW model of cosmology, given the 5D FLRW metric 

\begin{equation}
    dS^2 = -dt^2 + a^2(t)[dx^2+dy^2+dz^2+d\zeta^2]
\end{equation}

The EGB field equations in (3) reduce to the following two equations

\begin{align}
    k\rho=6H^2[1+2\alpha H^2]\\
    \frac{\ddot{a}}{a}=-\frac{1}{3}\frac{kP-3H^2}{1+4\alpha H^2}
\end{align}

Where, $H=\dot{a}/a$. We also have the continuity equation

\begin{equation}
    \dot{\rho}+4H(\rho + P)=0
\end{equation}

Eqs. (6) and (8) together describe the evolution of a homogeneous and isotropic five-dimensional universe including Gauss–Bonnet curvature corrections. Higher-order curvature effects come with the modification provided by the term proportional to the Gauss–Bonnet coupling $\alpha$.\\

In the limit when $\alpha \rightarrow 0$, one immediately recovers the standard 5D Einstein equations, for which it holds that $H^2 \propto \rho$. Equation (7) describes the acceleration of the cosmic expansion, modified by the denominator $1+4\alpha H^2$, which also contains the higher-order curvature terms. On the other hand, the continuity equation (8) is unaffected by the Gauss–Bonnet term, which reflects local energy-momentum conservation independence of the gravitational dynamics.\\

Equation (8) can be put in the form

\begin{equation}
\dot{H} = -2(1+\omega)H^2 \frac{1+2\alpha H^2}{1+4\alpha H^2}
\end{equation} 

Knowing that $P = \omega \rho$\\

From equation (6), we can write the relation between $H$ and $\rho$ in terms of the following roots

\begin{equation}
    H = \pm \sqrt{\frac{-1 \pm \sqrt{1+\frac{4}{3}k\alpha \rho}}{4\alpha}}
\end{equation}

We can write the two roots that correspond to a universe with $H<2(|\alpha|)^{-1/2}$ as follows

\begin{equation}
    H_{\pm} = \pm \frac{1}{2 \sqrt{|\alpha|}}\sqrt{1\mp \sqrt{1-\frac{4}{3}|\alpha|k\rho}}
\end{equation}

These two branches are the ones that remain real and physically relevant, such that the argument of the inner square root is non-negative, i.e. $1-\frac{4}{3}|\alpha|k\rho \geq 0$. This solution shows that the Hubble rate is always less than or equal to the Planck energy scale and the density has a maximum value $\rho_{max} \sim \rho_{planck}$ \\

Now, using the continuity equation, we can express the density as a function of the scale factor: 

\begin{equation}
    \rho(a) = C a^{-4\tilde{\omega}}
\end{equation}

where $\tilde{\omega} = 1+\omega$ and $C>0$\\

To get a better picture of the solution, we can rescale, such that $t \rightarrow \tau$ and $a \rightarrow \eta$, so that we can write the Hubble rate as

\begin{equation}
    h(\tau) = \frac{\eta'}{\eta} =  \pm \sqrt{1 \mp \sqrt{1- \eta^{-4\tilde{w}}}} = 2\sqrt{|\alpha|}H
\end{equation}

Where $\tau \equiv \frac{t}{2\sqrt{|\alpha|}}$, $\eta \equiv \frac{a}{a_c}$, $a_{c}^{4\tilde{\omega}} \equiv \frac{4}{3} \alpha k C$, and $'$ stands for the derivative with respect to $\tau$. In these rescaled variables, the dynamics is governed by a dimensionless Hubble parameter $h(\tau)$, and the Gauss–Bonnet corrections are absorbed into the choice of the characteristic scale $a_c$ and the rescaled time $\tau$.\\ 

From equation (9), we can also write an equation to represent the evolution of $h(\tau)$ 

\begin{equation}
    h'(\tau) = -(1+w)h^2\frac{2-h^2}{1-h^2}
\end{equation}

We notice that at $\eta = 1$, $h = \pm 1$, and $h'$ blows up, which causes a curvature singularity, where the Ricci scalar is proportional to $h'$.\\

\begin{figure}
    \centering
    \includegraphics[width=0.5\linewidth]{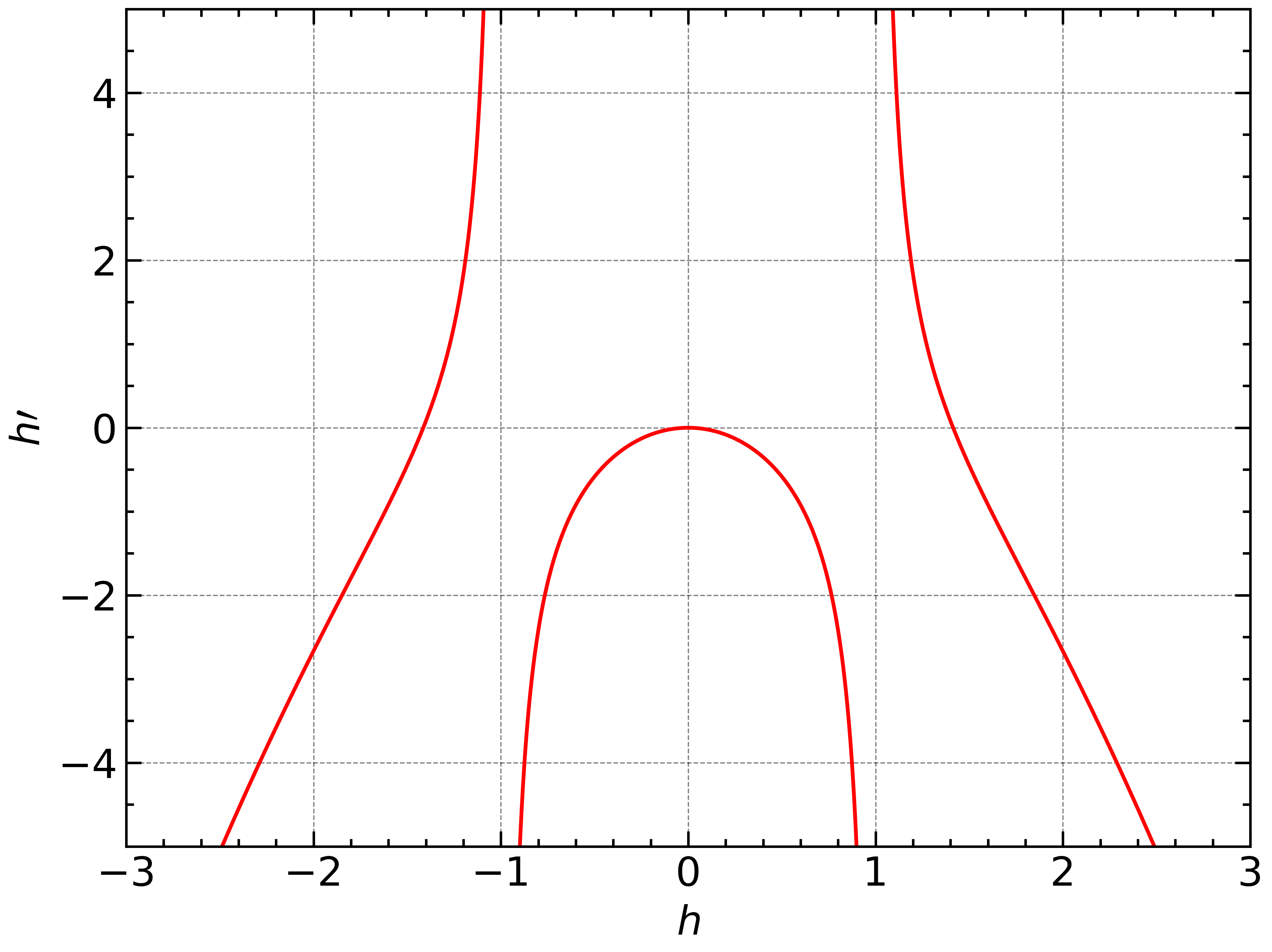}
    \caption{Phase-Space Diagram for the Hubble rate}
    \label{fig:Phasediagram}
\end{figure}
 
To get a better intuition for the solution and this evolution one could look at the Phase diagram in fig(1), We can do similar analysis done in \cite{Awad_2013} to study equation (14), which has a fixed point at $h=0$, but that would not be useful for us here, where we are interested in analyzing the solution beyond $h=1$. As we see from the figure, $h'$ blows up as it approaches that point.\\

This dynamical system can also be further explored by treating it as a mechanical system, which will enable us to understand the nature of this singular behavior better.

\subsection{A Mechanical Analogue}
Combining equations (11) and (12), we get the following energy conservation equation for our mechanical system 

\begin{equation}
    E_k + E_p = \frac{1}{2}\eta'^2 - \frac{1}{2}\eta^2 h^2(\tau) = \frac{1}{2}\eta'^2 - \frac{1}{2}\eta^2(1 \mp \sqrt{1-\eta^{-4\tilde{\omega}}}) = 0
\end{equation}

And from (14), we can obtain the equation of motion for this system

\begin{equation}
    \eta'' = \eta \frac{(\sqrt{1-\eta^{-4\tilde{\omega}}} + (-1+\tilde{\omega})\eta^{-4\tilde{\omega}}\mp 1)}{\sqrt{1-\eta^{-4\tilde{\omega}}}}
\end{equation}

We notice that there are two branches for this system, which provide us with two potential energies and two solution classes given by

\begin{equation}
    V_\mp (\eta) = -\frac{1}{2}\eta^2h^2(\tau) = -\frac{1}{2}\eta^2(1 \mp \sqrt{1-\eta^{-4\tilde{\omega}}})
\end{equation}

Now, using equation (14) for $h'(\tau)$ we can get a solution for the time $\tau$ as a function of $h$

\begin{equation}
    \tau + c_1 =- \frac{1}{2}\frac{(1+\omega)}{h} + \frac{(1+\omega)}{2\sqrt{2}}tanh^{-1}(\frac{h}{\sqrt{2}})
\end{equation}

Looking at the positive sign potential we see that it corresponds to $H \geq \frac{1}{2\sqrt{|\alpha|}}$ which gives us a curvature that is greater that or equal to $1/l_s^2$, where $l_s$ is string length \cite{KUIROUKIDIS_2006}, which is outside the scope of our classical treatment and would require a full theory of quantum gravity where quantum effects can't be ignored at such scales. So, we consider only branches with a Hubble rate of $H \leq \frac{1}{2\sqrt{|\alpha|}}$. We also notice that when we let $w=0$, the solution becomes identical to the solution found in \cite{Awad_2016}, where quantum corrections due to Weyl anomaly generated higher-curvature terms, which left the initial singularity traversable. This shows that EGB higher-curvature terms indeed affect singularities in gravitational theories\\

It is also essential to notice that, when calculating the Ricci scalar using the expansion of $a$, $H$, and $\dot{H}$ around $t=0$, which will be shown in the following sections, we get that 

\begin{equation}
    R \sim t^{-\frac{1}{2}}
\end{equation}

Which compared to the big bang/crunch singularity with $(R \sim t^{-2})$, the singularity at hand is much weaker.\\

We can take the following notes about our singularity from the mechanical system; 

\begin{enumerate}
    \item Even though the acceleration blows up at $\tau = 0$, $\eta=1$ and $\eta' = 1$, from equation (15) we see that the system needs a finite amount of work to evolve from the singular point to any other point with a finite scale factor $\eta$. This differs from the Big Bang/Big Crunch singularity, which requires an infinite amount of work to evolve from the singularity; hence, it is a much weaker singularity. 
    \item At $\eta = 1$ and $\eta' = 1$, the system is bounded, where it can't go to values less than $\eta = 1$ and can't stay at $\eta' = 0$, or it will be in contradiction with the energy condition from equation (15). We can imagine that the system reverses its motion by bouncing back with the opposite velocity.
    \item By looking at equations (18) with the initial conditions $\tau = 0$ at $\eta = 1$, we see from fig(2) that there are two disjoint regions for $|h| \leq 1$ with $\tau \geq 0$ and $\tau < 0$, since in the region with $\tau < 0$ the solution can't reach the other half with $\tau \geq 0$ where $\Delta \tau = \tau|_{h=1} - \tau|_{h=-1} \neq 0$.
\end{enumerate}

  \begin{figure}
      \centering
      \includegraphics[width=0.5\linewidth]{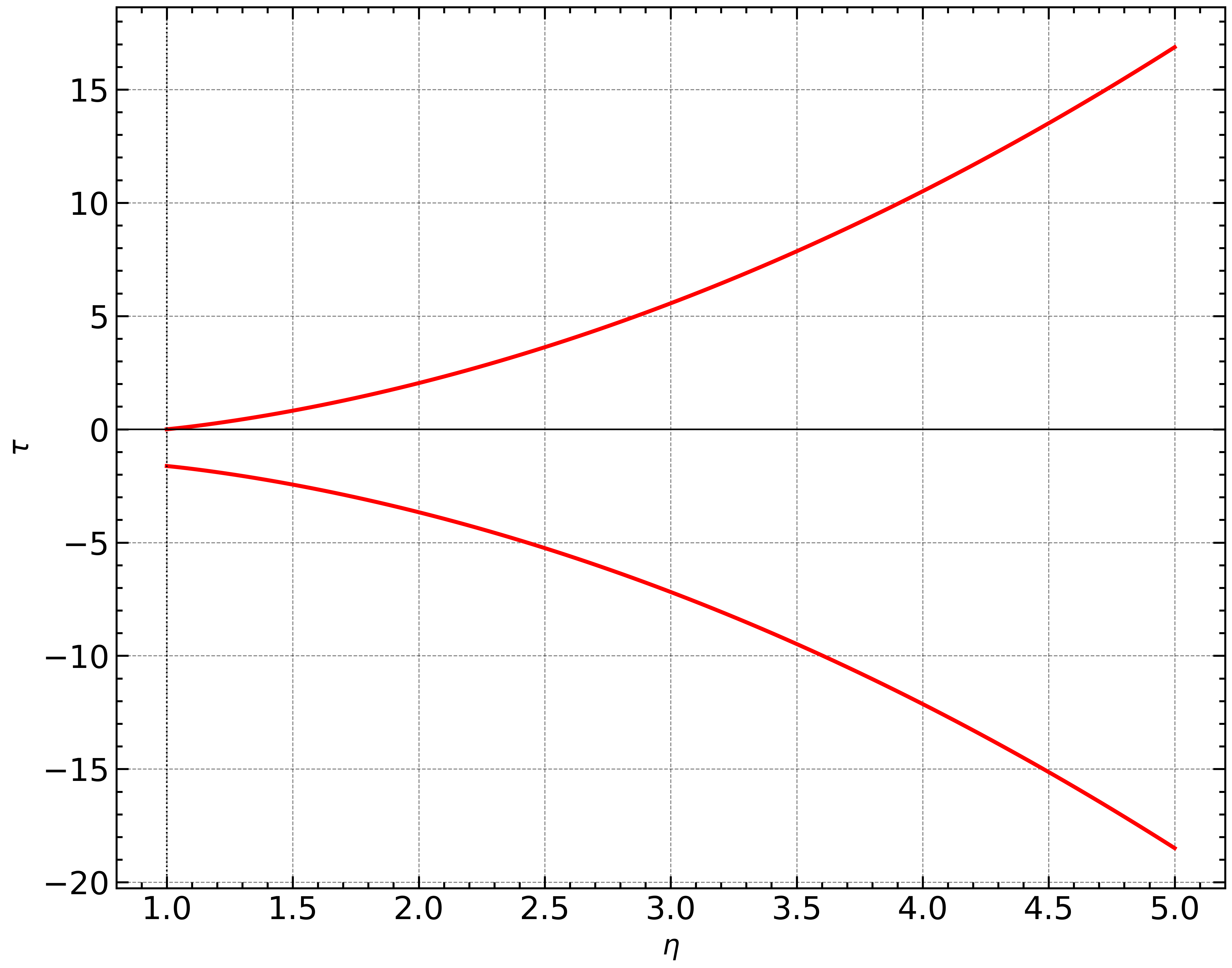}
      \caption{Time $\tau$ as a function of Scale Factor $\eta$, showing the two disjoint solution branches with $\tau \geq 0$ and $\tau < 0$.}
      \label{fig:placeholder}
  \end{figure}

So, the spacetime consists of two disjoint manifolds, preventing it from bouncing back with an opposite sign Hubble rate, where it is discontinuous at the singularity. This motivates us to show that we can glue the two manifolds together using appropriate junction conditions, thereby obtaining a geodesically complete bouncing solution, and to assess the strength of the singularity to ensure that a timelike object will not be destroyed along that geodesic. 

\subsection{Geodesics Extension and Strength of Singularity}
In this section, we are going to show a $C^2$ geodesic extension to non-spacelike geodesics beyond the singular point by joining the two solution branches, and then we will show that this singularity is weak according to Tipler and Krolak singularity strength conditions.\\

Starting with a test particle Lagrangian in the 5D FLRW spacetime

\begin{equation}
    \mathcal{L} = \frac{1}{2} \left( -\left(\frac{dt}{d\lambda}\right)^2 + a(t)^2 \sum_{i=1}^4 \left(\frac{dx^i}{d\lambda}\right)^2 \right)
\end{equation}

where $i=1, 2, 3, 4$

The conjugate momenta for the spatial coordinates $x^i$ are

\begin{equation}
    p_i = \frac{\partial \mathcal{L}}{\partial \dot{x}^i} = a(t)^2 \frac{dx^i}{d\lambda} = v^i
\end{equation}

and so we can write, 

\begin{equation}
    \frac{dx^i}{d\lambda} = \frac{v^i}{a^2} = f^i(\lambda)
\end{equation}

We can also use the constraint 

\begin{equation}
    -\left(\frac{dt}{d\lambda}\right)^2 + a(t)^2 \sum_{i=1}^4 \left(\frac{dx^i}{d\lambda}\right)^2 = -s
\end{equation}

And so we get 

\begin{equation}
    \frac{dt}{d\lambda} = \pm \sqrt{s + \frac{v^2}{a^2}} = g(t)
\end{equation}

Where  $s=-u^\mu u_\mu$, such that $u^\mu$ is the tangent vector, $\lambda$ is a non-spacelike affine parameter and the parameter $s$ controls the type pf $\lambda$, where if $s=1$, $\lambda$ is a timelike affine parameter, and if $s=0$, then $\lambda$ is a null affine parameter.\\

It is important to note that $f^i$ and $g$ are continuous in $\lambda$ and Lipschitz continuous in $t$, so, according to the Picard-Lindelöf theorem, there exist unique solutions for the geodesic equations. Still, since the scale factor is not continuous at the singularity, the geodesics end at that point, on the other hand we can define a new region of spacetime beyond this point, where as we will see later in this section, is that since the scale factor and Hubble rate are finite at the singular point, so it would be a valid extension, where the singularity would be classified as a sudden singularity, which is known to be geodesically extendable \cite{Barrow_2013, PhysRevD.74.064030, Nojiri_2005}

We begin by writing an expansion of $\eta$ around $\tau=0$ using equation (18) and (13)

\begin{equation}
    \eta(\tau) = 1+ |\tau| - \frac{2}{3}|\tau|^\frac{3}{2} + O(\tau^2)
\end{equation}

and using the definition of $\eta$ and $\tau$, to get a scale factor for the extended spacetime joining the two regions for $t \geq\ 0$ and $t < 0$ 

\begin{equation}
    a(t) = a_0 [1+|H_0t|-\frac{2}{3}|H_0t|^\frac{3}{2}] + O(t^2)
\end{equation}
where $H_0 = H(0) = \frac{1}{2\sqrt{|\alpha|}}$ and 

\begin{equation}
    H(t) = H_0[1- (H_0t)^\frac{1}{2}] + O(t^2)
\end{equation}

Integrating equation (24) with the initial condition of $\lambda = 0$ at $t=0$ we get 

\begin{equation}
    t(\lambda) =\chi \lambda - sign(\lambda)\frac{H_0v^2}{2a_0^2}\lambda^2 + O(t^3)
\end{equation}
where $\chi = (s+ \frac{v^2}{a_0^2})^2$, this enables us to write the scale factor in terms of the affine parameter as follows
\begin{equation}
    a(\lambda) = a_0[1+\chi H_0\lambda] + O(\lambda^3)
\end{equation}

Integrating equation (22) with the initial condition of $\lambda = 0$ at $x^i = x_0^i$ we get 

\begin{equation}
    x^i(\lambda) = x_0^i + \frac{v^i}{a_0^2}\lambda + sign(\lambda)\frac{H_0v^2}{a_0^2}\chi \lambda^2 +O(\lambda^3)
\end{equation}

Now we have $C^2$ geodesics for positive and negative values of $\lambda$, defining geodesic completeness. Still, we need to demonstrate the validity of this extension by assessing singularity strength using the criteria introduced by Tipler \cite{TIPLER19778} and Krolak \cite{CLARKE1985127}.

\subsection*{The singularity strength}
Starting with the lightlike geodesics, which reach a Tipler strong singularity at $\lambda = \lambda_0$ iff 

\begin{equation}
   \lim_{\lambda \to \lambda_0} \int_0^\lambda d\lambda' \int_0^{\lambda'} d\lambda'' R_{ab} u^a u^b, \quad \text{diverges}
\end{equation}

Using the above expansions, we get that 

\begin{equation}
    \lim_{\lambda \to 0} \int_0^\lambda d\lambda' \int_0^{\lambda'} d\lambda'' R_{ab} u^a u^b = C \lim_{\lambda \to 0} \lambda^{3/2} = 0
\end{equation}\\

It reaches Krolak strong singularity at $\lambda = \lambda_0$ iff

\begin{equation}
    \lim_{\lambda \to \lambda_0} \int_0^\lambda d\lambda' R_{ab} u^a u^b, \quad \text{diverges}
\end{equation}

we get 

\begin{equation}
    \lim_{\lambda \to 0} \int_0^\lambda d\lambda' R_{ab} u^a u^b = C' \lim_{\lambda \to 0} \lambda^{1/2} = 0
\end{equation}\\

But for the timelike geodesics to reach Tipler strong singularity, it is necessary that 

\begin{equation}
    \lim_{\lambda \to \lambda_0} \int_0^\lambda d\lambda' \int_0^{\lambda'} d\lambda'' |R^i_{ajb} u^a u^b|, \quad \text{diverges}
\end{equation}

we get 

\begin{equation}
    \lim_{\lambda \to \lambda_0} \int_0^\lambda d\lambda' \int_0^{\lambda'} d\lambda'' |R^i_{ajb} u^a u^b| = C_1 \lim_{\lambda \to 0} \lambda^{3/2} = 0
\end{equation}\\

and for the timelike geodesics to reach Krolak strong singularity, it is necessary for 

\begin{equation}
    \lim_{\lambda \to \lambda_0} \int_0^\lambda d\lambda' |R^i_{ajb} u^a u^b|, \quad \text{diverges}
\end{equation}

and we get 

\begin{equation}
    \lim_{\lambda \to \lambda_0} \int_0^\lambda d\lambda' |R^i_{ajb} u^a u^b| = C_2 \lim_{\lambda \to 0} \lambda^{1/2} = 0
\end{equation}\\

According to Tipler and Krolak, the singularity is weak; hence, it cannot destroy a physical object indefinitely, which supports our extension that renders the spacetime geodesically complete. However, our work is not complete; we must demonstrate that our extended spacetime satisfies the field equations. We will do so by verifying the extension against the junction conditions for Gauss-Bonnet gravity, as discussed in the following section.

\subsection{Junction conditions for 5D FLRW with the Gauss-Bonnet corrections}

In this section, we calculate the junction conditions for our extended solution using the general junction conditions for the 5D Gauss-Bonnet gravity in \cite{dolezel2000junctionconditionsgravitytheories}, starting with some definitions.

\begin{equation}
    ds^2 = \epsilon dw^2 + \gamma_{ij} dx^i dx^j
\end{equation}

Where this is the line element of the metric (g) in the Gaussian normal coordinates, and the constant $w$ 4D hypersurface $\Sigma$ with metric $\gamma_{ij}$ and $(i = 1, 2, 3, 4)$, we also have $n^\mu$ which is the normal vector to the family of hypersurfaces with $n^\mu n_\mu = \epsilon = 1 \ or\ -1$ for spacelike or timelike hypersurface, respectively. \\

The extrinsic curvature is given by 

\begin{equation}
    K_{ij} = -\frac{1}{2} \gamma_{ij,w}
\end{equation}

We can write the field equations in (3) in terms of intrinsic and extrinsic curvature of the 4D-hypersurface as derived in \cite{dolezel2000junctionconditionsgravitytheories}, starting with the Einstein tensor

\begin{align}
G^w_w &= -\frac{1}{2} {}^4 R + \frac{1}{2} \epsilon [K^2 - Tr(K^2)] , \\
G^w_i &= \epsilon [K_{|i} - K^j{}_{i|j}] , \\
G^i_j &= {}^4 G^i_j + \epsilon \left[ K^i_{j ,w} - \delta^i_j K_,w \right] + \epsilon \left[ -K K^i_j + \frac{1}{2} \delta^i_j Tr(K^2) + \frac{1}{2} \delta^i_j K^2 \right] .
\end{align}

We use this to join the two spacetimes (which are solutions of the field equations) at the hypersurface with $w=0$ as follows

\begin{align}
\lim_{\sigma \to 0} \int_{-\sigma}^{\sigma} G^w_w dw &= [\mathbf{G}^w_w] = 0 \\
\lim_{\sigma \to 0} \int_{-\sigma}^{\sigma} G^w_i dw &= [\mathbf{G}^w_i] = 0 \\
\lim_{\sigma \to 0} \int_{-\sigma}^{\sigma} G^i_j dw &= [\mathbf{G}^i_j] = \epsilon ( [K^i_j] - \delta^i_j [K] )
\end{align}

Where we define $[A] = A(0^+) - A(0^-) = A^+ -A^-$ to be the jump in "A" across the junction $w=0$. We notice that the contributing terms in the limit $\sigma \rightarrow 0$ are only the terms $\sim(K^i_j,_w)$.\\

Now, for the Gauss-Bonnet term, there exist no linear terms in $(K^i_j,_w)$ in $H^w_w$ nor in $H^w_i$, so they don't contribute. However, the component $H^i_j$ is different and from \cite{dolezel2000junctionconditionsgravitytheories} we can write it as follows

\begin{align}
H^i_j &= \left\{ K^i_j,_w \right\} (2 Tr(K^2) - 2K^2) + \left\{ K^i_l,_w \right\} (4 K K^l_j - 4 K_{jn} K^{nl}) \nonumber \\
&\quad + \left\{ K_{il,w} \right\} (4 K K^{li} - 4 K^i_n K^{nl}) + \left\{ K_{,w} \right\} (4 K^i_n K^n_j - 4 K K^i_j) \nonumber \\
&\quad + \left\{ K_{mn,w} \right\} (4 K^i_j K^{mn} - 4 K^{mi} K^n_j) \nonumber \\
&\quad + \left\{ K_{mn,w} \right\} \delta^i_j (4 K^m_p K^{pn} - 4 K K^{mn}) + \left\{ K_{,w} \right\} \delta^i_j (2 K^2 - 2 Tr(K^2)) \nonumber \\
&\quad + \epsilon \left( -4 {}^4 R^i_{mjn} K^m_{n,w} - 4 {}^4 R^m_j K^i_{m,w} - 4 {}^4 R^{mi} K_{jm,w} \right) \nonumber \\
&\quad + \epsilon \left( 4 {}^4 R^i_j K_{,w} + 2 {}^4 R K^i_{j,w} + 4 \delta^i_j {}^4 R^{mn} K_{mn,w} - 2 \delta^i_j {}^4 R K_{,w} \right) \nonumber \\
&\quad + \dots
\end{align}

where $(+\dots)$ are terms of zeroth order in $K^i_{j,w}$, which disappear as $\sigma \rightarrow 0$, and $Tr(K^2) \equiv K^{ij}K_{ij}$.\\

Now, let's calculate these junction conditions for our extended spacetime

\begin{equation}
    a(t) = a_0[1 + |H_0t| - \frac{2}{3}|H_0t|^\frac{3}{2}]+O(t^2)
\end{equation}

Also, we have the Hubble rate and its derivative; 

\begin{align}
H(t) &= sgn(t) H_0 - sgn(t) H_0^{3/2} \sqrt{|t|} + O(t) \\
\dot{H}(t) &= 2\delta(t) H_0 - \delta(t) H_0^{3/2} \sqrt{|t|} - \frac{H_0^{3/2}}{\sqrt{|t|}} + O(t)
\end{align}

We join the two manifolds for all values of $(t)$ and calculate the jumps in the stress tensor components $[\mathbf{T^i {}_j}]$, the hypersurface is spacelike, where $w = t$, $\epsilon = +1$ and $\gamma_{ij} = a^2(t)\delta_{ij}$, so the extrinsic curvature is given by
\begin{equation}
    K_{ij} = -\frac{1}{2}\gamma_{ij,t} = -\dot{a}a\delta_{ij} = -a^2H\delta_{ij}
\end{equation}

and 

\begin{equation}
    K = -4H, \quad Tr(K^2) = 4H^2 
\end{equation}

Using equations (47), (51), (52), and the fact that ${}^4R = 0$, since the hypersurface is spatially flat, we get 

\begin{equation}
    H^i_j = 48H^4\delta^i{}_j - 12\dot{H}H^2\delta^i{}_j
\end{equation}

Since 

\begin{equation}
    \kappa[\mathbf{T^i_j}] = [\mathbf{G^i{}_j}] + [\mathbf{H^i{}_j}] \quad and \quad [\mathbf{H^i{}_j}] = \lim_{\sigma \to 0} \int_{-\sigma}^{\sigma} H^i_j dw
\end{equation}

From equations (46), (53), and (54), we get the following 

\begin{align}
    \kappa[\mathbf{T^i_j}] &= [K^i_j] - \delta^i_j [K] + [\mathbf{H^i{}_j}] \nonumber\\
    &= 6H_0[1 - 4H^2(0)]\delta^i_j \nonumber\\
    &= 6H_0  \delta^i_j = \mathbf{S^i_j}
\end{align}

Here $\mathbf{S^i{}_j}$ is the surface stress energy tensor, and now we can write the junction conditions for joining the two manifolds as follows

\begin{align}
\kappa [\mathbf{T}^w_w] &= [\mathbf{G}^w_w] + [\mathbf{H}^w_w] = 0, \\
\kappa [\mathbf{T}^w_i] &= [\mathbf{G}^w_i] + [\mathbf{H}^w_i] = 0, \\
\kappa [\mathbf{T}^i_j] &= [\mathbf{G}^i_j] + [\mathbf{H}^i_j] =6H_0 \delta^i_j= \mathbf{S}^i_j
\end{align}

We see from equation (58) that $\mathbf{S_{ij}}= 6H_0a^2_0\delta_{ij}$, and so we get a delta function jump in pressure,

\begin{equation}
    P \rightarrow P + (6H_0 a^2_0)\delta(t)
\end{equation}

This makes sense, where we are extending the solution through a sudden singularity.

\section{The Black Hole Singularity}

In this part, we are going to look at the black hole singularity, specifically using the Boulware and Deser Black hole solution \cite{PhysRevLett.55.2656, GARRAFFO_2008}, with the following metric

\begin{equation}
    ds^2 = -f(r)dt^2 + f(r)^{-1}dr^2 + r^2d\Omega^2_3
\end{equation}

where $d\Omega^2_3$ is the metric of a unitary 3-sphere, and we can write the function $f(r)$ as follows

\begin{equation}
    f(r) = \frac{r^2 + 4\alpha \pm \sqrt{r^4 + 16\alpha m}}{4\alpha}
\end{equation}

The solution with the (-) sign is the physically stable solution, and the (+) solution is unstable, as has been shown by Boulware and Deser. Furthermore, the (-) sign solution goes to the Schwarzschild solution in the limit $\frac{16\alpha m}{r^4} \rightarrow 0$

\begin{equation}
    f(r) = 1 - \frac{2m}{r^2} 
\end{equation}

So, we are going to consider the (-) sign solution with $\alpha > 0$ as a physical and stable solution, in which case we have the following horizon 

\begin{equation}
    r_h = \sqrt{2}\sqrt{m - \alpha}
\end{equation}

and we find that the Ricci scalar curvature blows up at $r=0$ as 

\begin{equation}
    R \sim \sqrt{\frac{m}{\alpha}}\frac{1}{r^2}
\end{equation}

But in the same spirit as our analysis of the initial singularity, we will show that this blow-up of the Ricci curvature does not indicate geodesic incompleteness at the singular point of the black hole, which is the main characteristic of a spacetime physical singularity. 

Starting with the Lagrangian for a test particle,

\begin{equation}
    \mathcal{L} = -\frac{1}{2}[-f(r)\dot{t}^2 + f(r)^{-1}\dot{r}^2 + r^2(\dot{\theta}^2 + sin^2(\theta)\dot{\phi}^2 + cos^2(\theta)\dot{\psi}^2)]
\end{equation}

and setting $\theta = \pi/4$ for the radial geodesic in the equatorial plane, we get the following

\begin{equation}
    \dot{t} = \frac{E}{f(r)}, \quad \dot{\phi} = \frac{L_\phi}{r^2}, \quad \dot{\psi} = \frac{L_\psi}{r^2}, \quad\ and \quad L^2 = \frac{1}{2}(L_\phi^2 + L_\psi^2) 
\end{equation}

We also have the constraint

\begin{equation}
    -\frac{1}{2}[-f(r)\dot{t}^2 + f(r)^{-1}\dot{r}^2 + r^2(\dot{\theta}^2 + sin^2(\theta)\dot{\phi}^2 + cos^2(\theta)\dot{\psi}^2) = \delta
\end{equation}

where $\delta = 0$ for a null geodesic and $-1$ for a timelike geodesic, and so we get the following geodesic equation 

\begin{equation}
    \dot{r}^2 = E^2 - f(r)(\frac{L^2}{r^2} + \delta)
\end{equation}

with an effective potential 

\begin{equation}
    V(r) = f(r)(\frac{L^2}{r^2} + \delta)
\end{equation}

\begin{figure}
    \centering
\begin{minipage}[t]{.5\textwidth}
    \centering
    \includegraphics[width=0.9\linewidth]{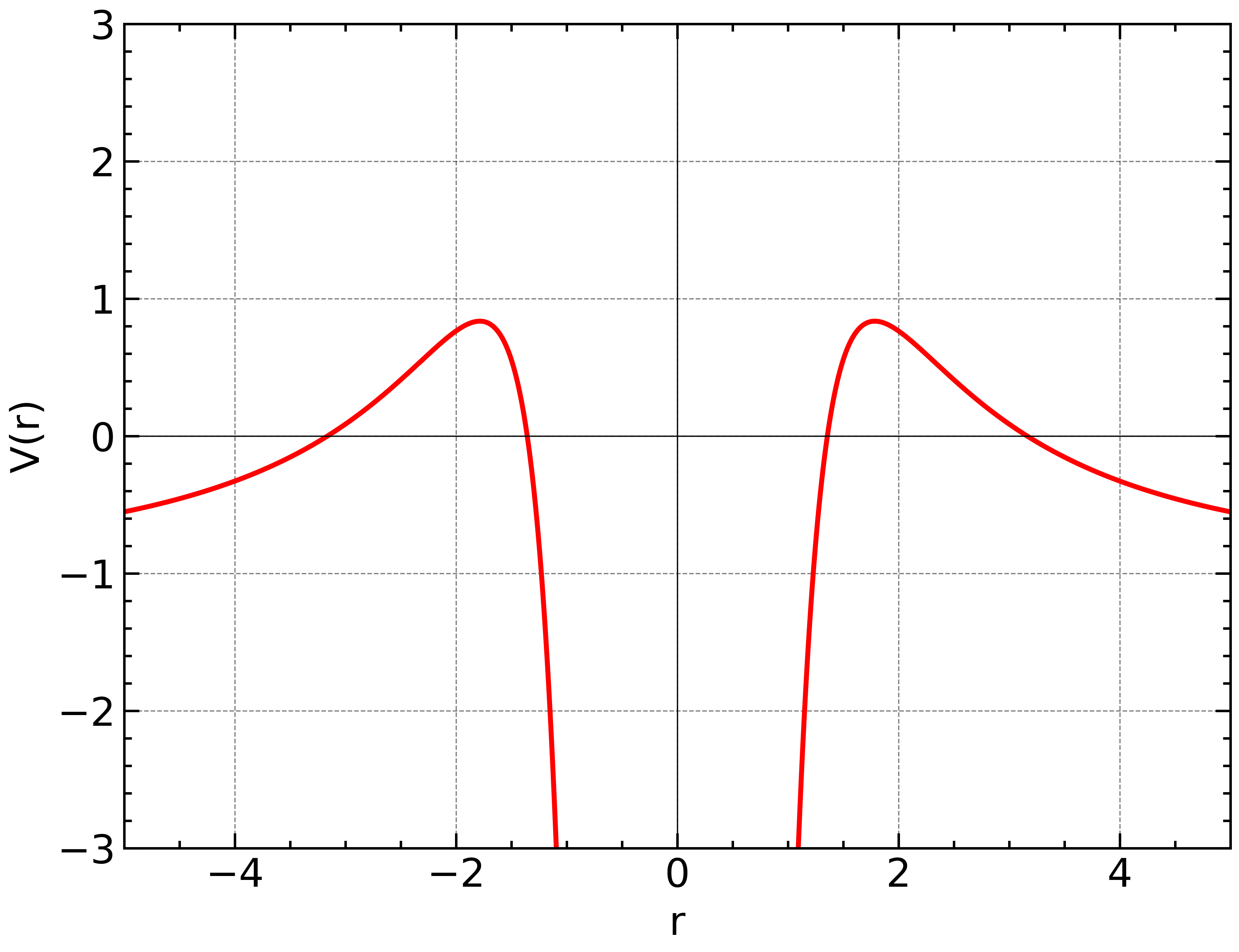}
    \captionof{figure}{$V(r)$ for a timelike object in equation (69), with $m=1$, $\alpha = 0.08$ and $L^2 = 10$ }
    \label{fig:placeholder}
\end{minipage}%
\begin{minipage}[t]{.5\textwidth}
     \centering
    \includegraphics[width=0.9\linewidth]{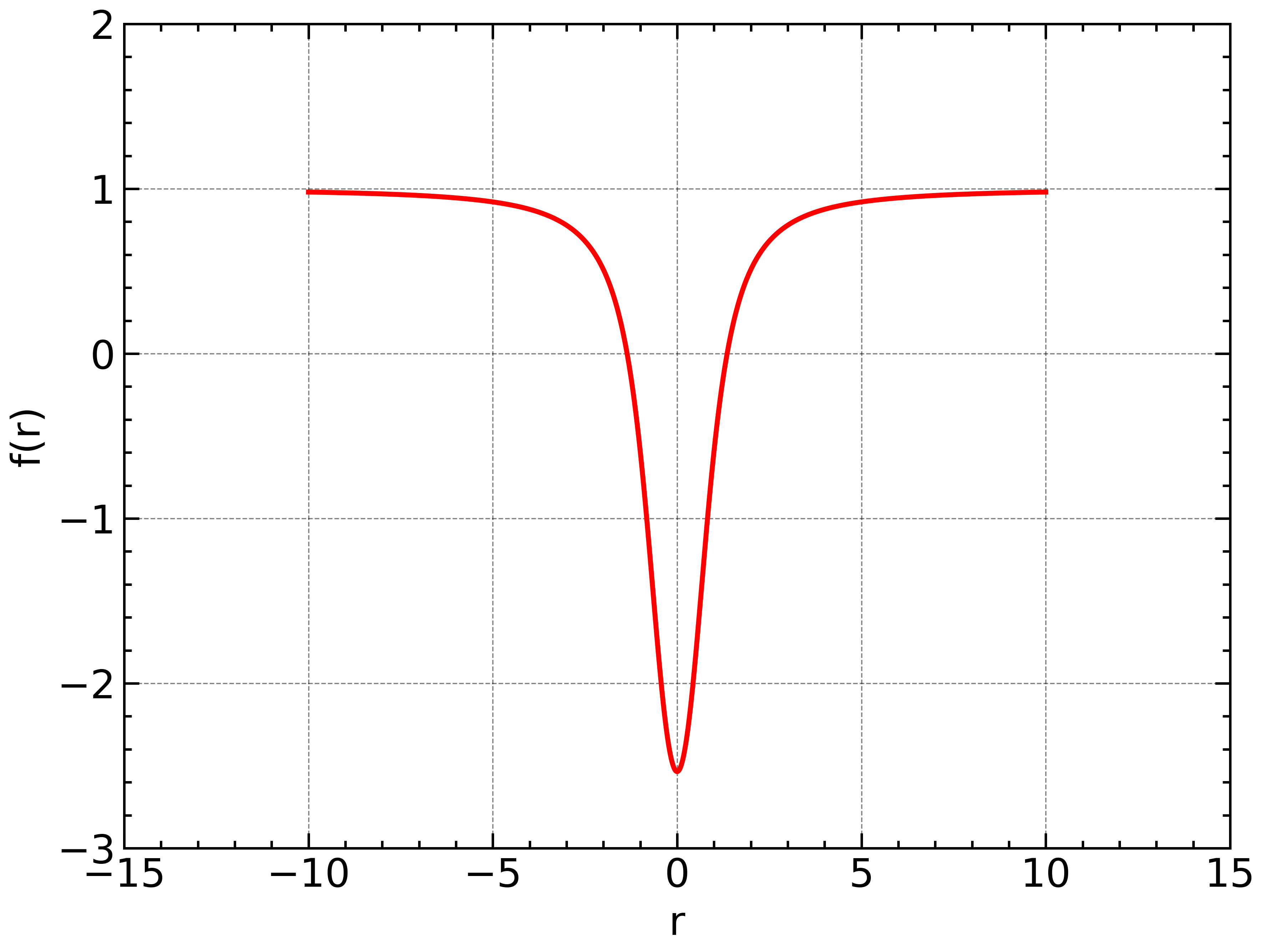}
    \captionof{figure}{The negative sign f(r) with $m=1$, $\alpha = 0.08$}
    \label{fig:placeholder}
\end{minipage}
\end{figure}

We immediately notice that the geodesic is indeed singular at $r=0$, and it is also shown from the potential in fig(3). But that singularity is due to the coupling of the angular momentum (ie, the centrifugal force), and not inherent in the geometric nature of the point $r=0$ of the geodesic itself.\\

So, looking at the purely radial geodesic with $L=0$, we have 

\begin{equation}
     \dot{r}^2 = E^2 - f(r)\delta
\end{equation}

with an effective potential 

\begin{equation}
    V(r) = f(r)\delta
\end{equation}

This is not singular at $r=0$ as seen in fig(5) and fig(6). Due to the well-behaved function $f(r)$ shown in fig(4), it is essential to keep in mind that this is not a correction of our own making. It is a case of a geodesic with no angular momentum, that we use to test the true nature of the singular point of $r=0$ where it is unubstructed by the cetrifugal force, in fact it was proven in \cite{Nolan_2000} that all central singularities are strong in the case of non-radial geodesics and in \cite{Arrechea_2025} it was shown that conditions like Krolak and Tipler's used earliear for the initial singularity are used to determine the strength of singularities in the radial geodesics case, and so we will do precisely that. 

\begin{figure}
    \centering
\begin{minipage}[t]{.5\textwidth}
    \centering
    \includegraphics[width=0.9\linewidth]{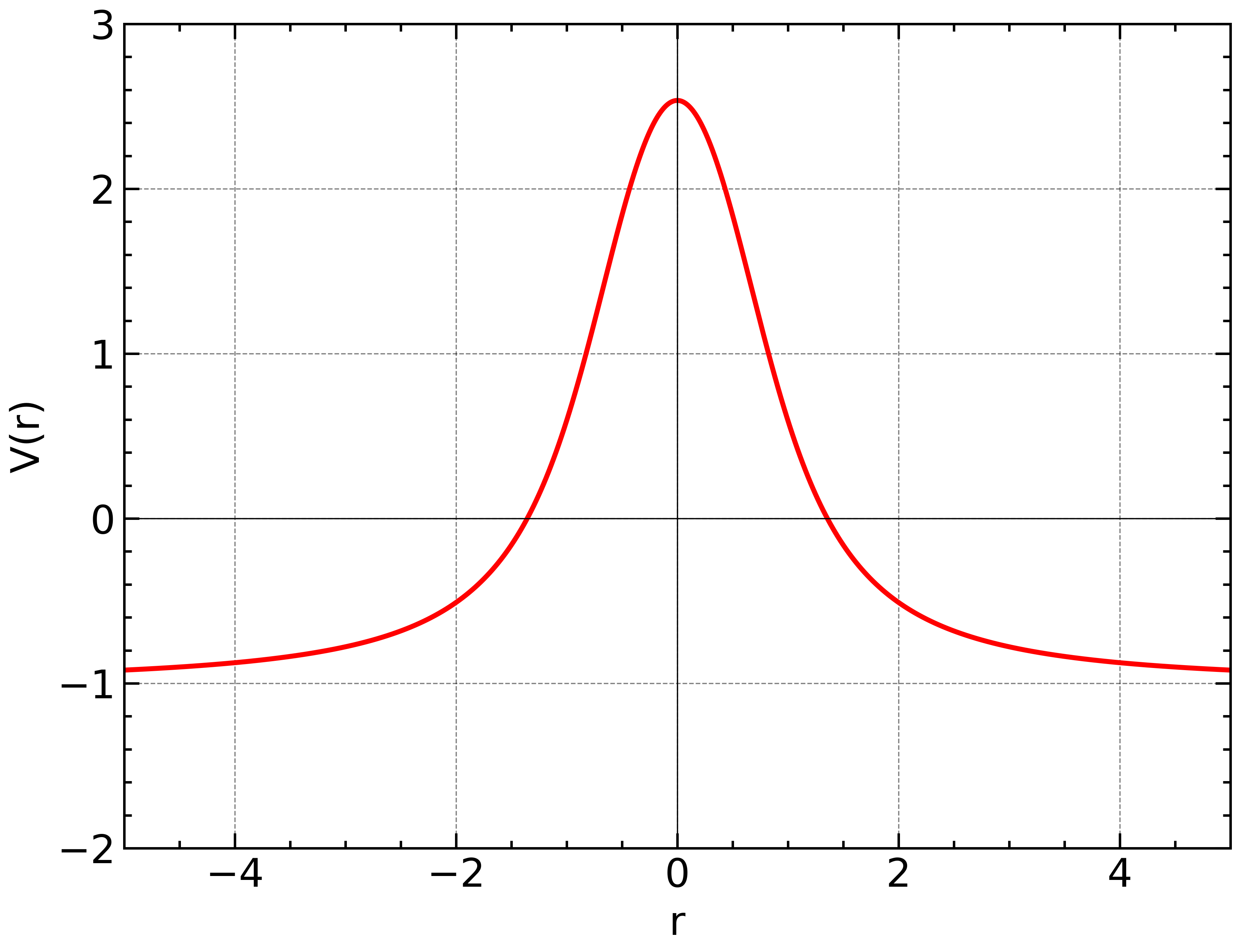}
    \captionof{figure}{$V(r)$ for a timelike object in equation (71), with $m=1$, $\alpha = 0.08$ and $L = 0$ }
    \label{fig:placeholder}
\end{minipage}%
\begin{minipage}[t]{.5\textwidth}
     \centering
    \includegraphics[width=0.9\linewidth]{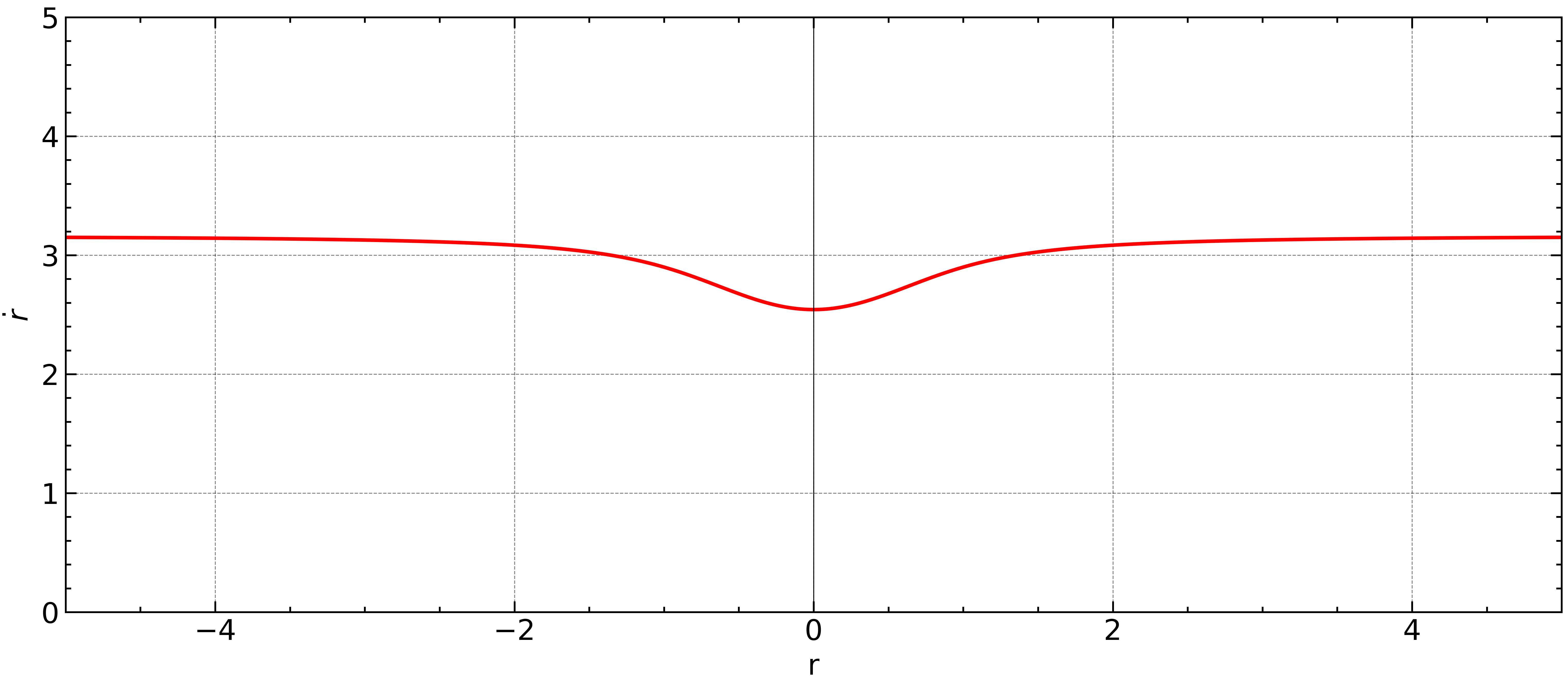}
    \captionof{figure}{Timelike $\dot{r}$ plote across $r=0$ with $E=3$, $m=1$, $\alpha = 0.08$, and $L=0$}
    \label{fig:placeholder}
\end{minipage}
\end{figure}

\subsection*{The singularity strength and geodesics extension}

We now test the strength of the singular point at $r=0$ using the purely radial geodesic, applying the Krolak and Tipler criteria for singularity strength. First, we define the following tangent vector
\begin{equation}
    U^a = (\dot{t}, \dot{r}, 0, 0, 0)
\end{equation}

where $\dot{t}$ and $\dot{r}$ are defined in equations (66) and (70) respectively. Expanding the function $f(r)$ and its derivatives for a very small $r$ as follows

\begin{align}
    f(r) &= \left( 1 - \sqrt{\frac{m}{\alpha}} \right) + \frac{r^2}{4\alpha} - \frac{r^4}{32 \alpha^{3/2} \sqrt{m}} + O(r^6), \\
     f'(r) &= \frac{r}{2\alpha} - \frac{r^3}{4 \alpha^{3/2} \sqrt{m}} + O(r^5), \\
     f''(r) &= \frac{1}{2\alpha} - \frac{3 r^2}{4 \alpha^{3/2} \sqrt{m}} + O(r^4)
\end{align}

Using equations (72) to (75), we can apply the Krolak and Tipler tests to the singularity. Starting with the Krolak necessary condition for timelike geodesics in equation (37), we get 

\begin{equation}
     \lim_{\lambda \to \lambda_0} \int_0^\lambda d\lambda' |R^i_{ajb} u^a u^b| \approx C\frac{\lambda_0}{\alpha} \neq \infty
\end{equation}

The Tipler necessary condition for timelike geodesics in equation (35), we have 

\begin{equation}
    \lim_{\lambda \to \lambda_0} \int_0^\lambda d\lambda' \int_0^{\lambda'} d\lambda'' |R^i_{ajb} u^a u^b| \approx  C'\frac{\lambda_0^2}{\alpha} \neq \infty
\end{equation}

where $\lambda$ is the affine parameter and $\lambda_0$ is its value at the singularity

Similarly, for null geodesics, we have the Krolak condition in equation (33) 

\begin{equation}
     \lim_{\lambda \to \lambda_0} \int_0^\lambda d\lambda' R_{ab} u^a u^b \approx 0
\end{equation}

and the Tipler condition for null geodesics in equation (31)

\begin{equation}
     \lim_{\lambda \to \lambda_0} \int_0^\lambda d\lambda' \int_0^{\lambda'} d\lambda'' R_{ab} u^a u^b \approx 0
\end{equation}

We see that the singularity is weak and physical objects can get through it without getting crushed, and from equations (70) and (74), we get

\begin{equation}
    \frac{d^2r}{d\lambda^2} = -\frac{f'(r)}{2} \approx \frac{-r}{4\alpha}
\end{equation}

For small values of $r$, equation (80) is a harmonic oscillator differential equation with the following solution

\begin{equation}
    r(\lambda) = Acos(\vartheta\lambda + \varphi) 
\end{equation}
with $\vartheta = \frac{1}{2\sqrt{\alpha}}$ and $\varphi$ is the phase\\

\begin{figure}
    \centering
    \includegraphics[width=0.5\linewidth]{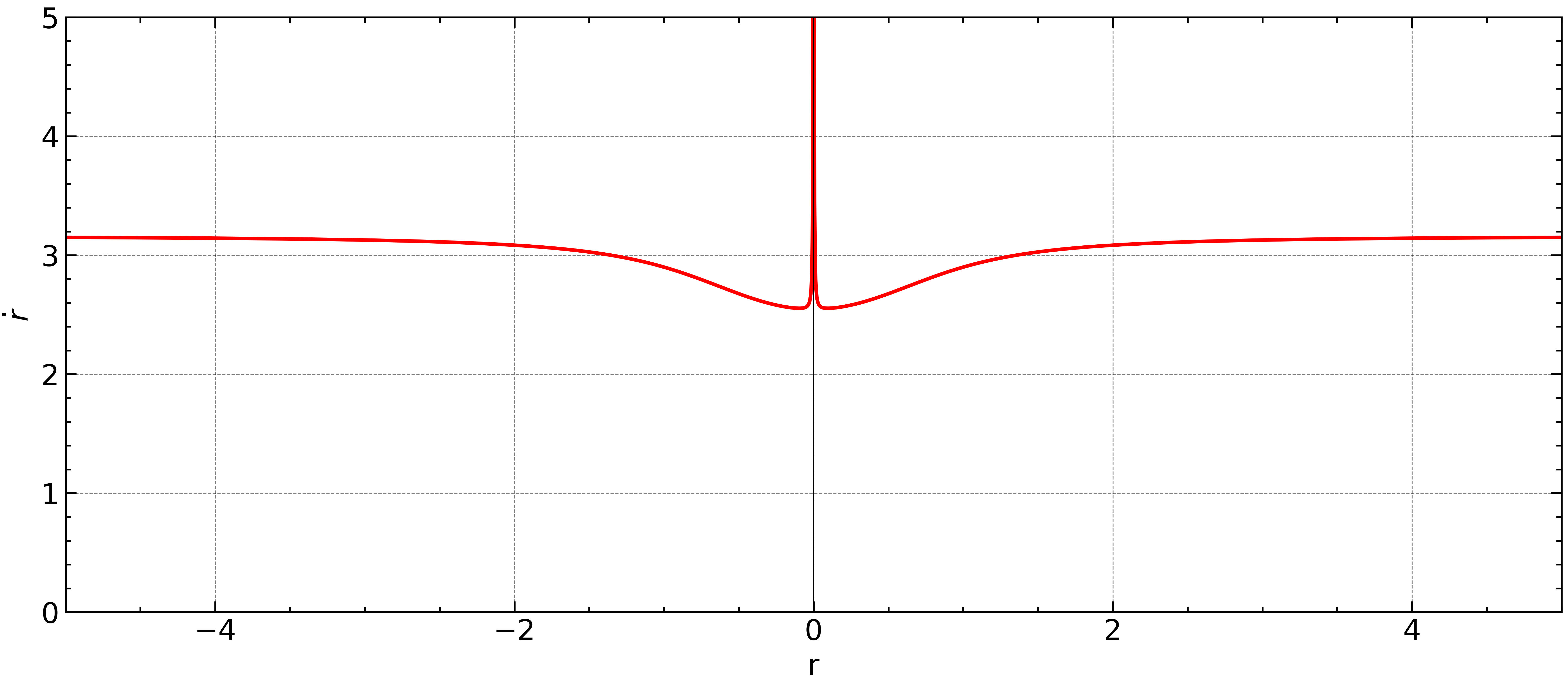}
    \caption{Timelike geodesic with small perturbation in L}
    \label{fig:placeholder}
\end{figure}

We find that by allowing our geodesics to be purely radial (ie. with vanishing angular momentum), there is no problem extending it beyond the singular point! Also, we found that the singularity, according to Tipler and Krolak, is weak in that case, but any small perturbation of the geodesics restores the singular behavior as shown in fig(7).\\

This behavior leads us to investigate the junction conditions between the two regions of the metric with $+r$ and $-r$, starting with the metric in the Gaussian normal coordinates, with $r \rightarrow 0$, $r$ being a timelike coordinate and $t$ a spacelike coordinate within the horizon

\begin{equation}
    ds^2 \approx -dw^2 + [(\sqrt{\frac{m}{\alpha}} -1)-\frac{(\sqrt{\frac{m}{\alpha}} -1)}{8\alpha}w^2]dt^2 + [(\sqrt{\frac{m}{\alpha}} -1)w^2 - \frac{(\sqrt{\frac{m}{\alpha}} -1)}{12\alpha}w^4]d\Omega^2_3
\end{equation}

Considering the 4D spacelike hypersurface with constant $w$ and using the leading order terms of the metric, we get the following extrinsic curvature

\begin{equation}
    K^t_t \sim 0, \quad K^i_j \sim -1/w \quad and \quad tr(K) \sim -3/w
\end{equation}

We find that the 3-sphere part of the metric leads to a diverging extrinsic curvature tensor, preventing the possibility of smooth junction conditions. This is a strange feature since in GR the metric function $f(r)$ itself is not well-behaved at $r=0$, while in EGB gravity $f(r)$ and its derivatives are finite! Furthermore, although this singularity is weak according to Tipler and Krolak, it is still a nontraversable one.

\section{Our Results and Singularity Theorems}
In this section, we would like to check our results against the Hawking-Penrose singularity theorems \cite{PhysRevLett.14.57, Hawking:1970zqf} since they work as important no-go theorems. Briefly, these theorems show that under certain energy conditions, such as the strong energy condition, the formation of trapped surfaces leads to geodesic (or acoustic) focusing as described by the Raychaudhuri equation, which indicates geodesic incompleteness. Furthermore, if the expansion parameter $\theta$, controlled by the Raychaudhuri equation, diverges, these geodesics are inextendible.

The Raychaudhuri equation for a geodesic congruence in $D$ dimensions \cite{PhysRev.98.1123, PhysRevD.98.024006} is given by
\begin{equation}
\frac{d\theta}{d\lambda} = -\frac{1}{D-2}\theta^{2} - \sigma_{\mu\nu}\sigma^{\mu\nu} + \omega_{\mu\nu}\omega^{\mu\nu} - R_{\mu\nu}k^{\mu}k^{\nu},
\end{equation}
where $\theta$ is the expansion parameter, $\sigma_{\mu\nu}$ is the shear, $\omega_{\mu\nu}$ is the twist (both vanish for hypersurface-orthogonal geodesics), and $R_{\mu\nu}$ is the Ricci tensor.

The key to applying these theorems in Gauss-Bonnet gravity is to realize that the "focusing condition" is not directly tied to the matter stress-energy tensor, but to the effective stress-energy tensor derived from the full field equations. The Gauss-Bonnet term contributes to this effective stress-energy tensor with a repulsive force at very high curvatures. Because of this repulsive effect, the effective stress-energy tensor can violate the Strong Energy Condition (SEC), even when the ordinary matter itself obeys it! This means that the "focusing condition" in the Raychaudhuri equation is not satisfied through the halting of the convergence of geodesics. This is clear from the calculation of the expansion $\theta $ $\sim$ $-H$, which never diverges but reaches a maximum value, in contrast to what happened in GR cosmology with divergent $H$ at the singularity. As we have seen in EGB cosmology, non-spacelike geodesics are extendable beyond the singular point, and the universe undergoes a contraction phase reaching a minimum size before it re-expands, rather than starting with a singularity in the past.

The 5D Boulware-Deser black hole presents a different picture, since the solution has an event horizon, i.e., a trapped surface forms, which is a key condition for singularity formation. For the effective Energy Condition, one can observe that as one approaches the center ($r \to 0$), the effective stress-energy tensor (derived from the GB term) satisfies a focusing condition as well see soon. The repulsive effects are not strong enough to overcome the gravitational pull.
As we shall see in the following, the presence of a trapped surface and the focusing of geodesics inevitably lead to geodesic incompleteness. The Gauss-Bonnet term, while altering the metric, does not resolve the central singularity.

\subsection{Raychaudhuri Equation for the Boulware-Deser Black Hole}

To see this closely, let us consider a congruence of ingoing radial null geodesics, with the following tangent vector $k^{\mu}$
\begin{equation}
k^{\mu} = \left(\frac{1}{f(r)}, -1, 0, 0, 0\right) 
\end{equation}
It satisfies the null condition $g_{\mu\nu}k^{\mu}k^{\nu} = 0$ and has the affine parameter $\lambda$ such that $k^{\mu}\nabla_{\mu}k^{\nu} = 0$. The expansion $\theta$ is defined as
\begin{equation}
\theta = \nabla_{\mu}k^{\mu}
\end{equation}
Calculating $\theta$ for the Boulware-Deser solution, one gets
\begin{equation}
\theta = \frac{-3}{r}.
\end{equation}
This measures the rate of change of the cross-sectional area of the congruence. From the tangent vector $k^{\mu} = dx^{\mu}/d\lambda$, we have
\begin{equation}
\frac{dr}{d\lambda} = k^{r} = -1,
\end{equation}
which leads to $\lambda \sim -r$, where we set the integration constant to zero, or $\lambda \to 0$ as $r \to 0$.
Notice that $\theta < 0$ for all $r > 0$, indicating that the congruence is converging as it moves inward, while at the event horizon $r = r_{h}$, $\theta$ is finite and non-zero, i.e., the horizon is a regular point.
But as $r \to 0^{+}$, $\theta \to -\infty$, which indicates convergence to a caustic at the center, signaling geodesic incompleteness and the presence of a singularity.
Another way to verify the calculation of $\theta$, one can start from the above Raychaudhuri equation with vanishing shear and twist, then calculate $R_{\mu\nu}k^{\mu}k^{\nu}$, which leads to a vanishing expression, then one gets 
\begin{equation}
\frac{d\theta}{d\lambda} = -\frac{1}{D-2}\theta^{2},
\end{equation}
which leads again to $\theta \sim \frac{3}{\lambda}$, confirming our previous result and the existence of inextendible geodesics.

This demonstrates that the Boulware-Deser black hole is geodesically incomplete, with the singularity at $r=0$ (or $\theta \to -\infty$) reached in finite affine parameter.

\section{Conclusion}
In this work, we have investigated how Einstein–Gauss–Bonnet (EGB) corrections alter the nature and strength of spacetime singularities in five-dimensional cosmological and black hole settings, with particular emphasis on geodesic completeness and the Tipler–Królak criteria. For the 5D FLRW cosmology, we found that Gauss–Bonnet terms replace the standard big bang/crunch singularity with a much weaker sudden singularity. At this point, the scale factor and Hubble rate remain finite while higher derivatives diverge, with the Ricci scalar behaving as $R \sim t^{-1/2}$ instead of the usual $R \sim t^{-2}$. Using a mechanical analogue, we showed that only a finite amount of work is needed to move away from the singular point, and that the phase space splits into two branches. By constructing an explicit $C^2$ extension of non-spacelike geodesics and imposing Gauss–Bonnet junction conditions, we obtained a geodesically complete bouncing spacetime with a well-defined surface stress–energy tensor corresponding to a delta-function jump in the pressure. The relevant Tipler and Królak integrals remain finite, confirming that this cosmological singularity is weak.\\

For the Boulware–Deser black hole, we showed that although curvature invariants diverge as $r\rightarrow 0$, this does not necessarily imply a strong singularity along all geodesics. Non-radial geodesics remain strongly singular due to the centrifugal barrier, consistent with previous results. However, along purely radial geodesics, the Gauss–Bonnet corrections soften the central singularity: the effective potential and the geodesic equations remain regular enough that the Tipler and Królak criteria indicate a weak singularity, with near-center motion approximated by a harmonic-oscillator-like behavior.\\

Our investigation of junction conditions in the black hole case shows that although higher-curvature corrections alter the nature of the singularity, geodesics are still inextendible since the extrinsic curvature tensor components diverge at the singularity. Furthermore, our results are consistent with the Penrose-Hawking singularity theorems since in Gauss-Bonne black holes, geodesics suffer from focusing, i.e., the expansion parameter diverges, while in cosmology, there is no focusing since the expansion parameter remains finite at the singularity, allowing for geodesic extendability.\\

Our analysis shows that Gauss–Bonnet higher-curvature terms clearly weaken both cosmological and black-hole singularities as a result of the repulsive effect near singularities. Although in EGB cosmology these effects were strong enough to render the singular region traversable, it was not strong enough to change the singular nature of the black hole central region. This approach could provide us with a viable mechanism for investigating singular solutions and can be extended to study various solutions. Specifically, those with less symmetry or additional parameters, e.g., rotating or charged black holes in Gauss–Bonnet gravity. These investigations may shed more light on how generic these singularity-weakening effects are.

\section{Aknowlegement}
The work of A.A. is partially supported by the Science, Technology \&\ Innovation Funding Authority (STDF) under grant number 50806.

\subsection*{References}


%

\end{document}